\documentclass[conference]{IEEEtran}

\ifCLASSINFOpdf
\else
\fi

\usepackage{url}

\usepackage[nolist]{acronym}
\usepackage{subfigure} 

\usepackage{color}
\usepackage{graphicx}
\begin{document}
%
\title{NAVI: Neighbor Aware Virtual Infrastructure for Information Dissemination in Vehicular Networks}

\author{\IEEEauthorblockN{
Pedro M. d'Orey\IEEEauthorrefmark{1},
Nitin Maslekar\IEEEauthorrefmark{1},
Idoia de la Iglesia\IEEEauthorrefmark{1} and 
Nikola K. Zahariev\IEEEauthorrefmark{1}}
\IEEEauthorblockA{\IEEEauthorrefmark{1} NEC Laboratories Europe, Kurf{\"u}rsten-Anlage 36, 69115 Heidelberg, Germany\\
\{pedro.dorey, nitin.maslekar, idoia.delaiglesia, nikola.zahariev\}@neclab.eu}
}


\maketitle

\begin{acronym}
	\acro{BTP}{Basis Transport Protocol}
	\acro{NARR}{Neighbors Above Range Ratio}
	\acro{CACC}{Cooperative Adaptive Cruise Control}
	\acro{CAM}{Cooperative Awareness Message}
	\acro{D2D}{Device to Device}
	\acro{DENM}{Decentralized Environmental Notification Message}
	\acro{DSRC}{Dedicated Short Range Communication}
	\acro{FOT}{Field Operational Test}
	\acro{GPS}{Global Positioning System}
	\acro{ITS}{Intelligent Transportation Systems}
	\acro{IRT}{Inter-Reception Time}
	\acro{LAN}{Local Area Network}
	\acro{LOS}{Line of Sight}
	\acro{NAR}{Neighborhood Awareness Ratio}
	\acro{NIR}{Neighborhood Interference Ratio}	
	\acro{NLOS}{Non Line of Sight}
	\acro{PDR}{Packet Delivery Ratio}
	\acro{RNARI}{Ratio of Neighbors Above Region of Interest}
	\acro{RNAR}{Ratio of Neighbors Above Range}
	\acro{RSU}{Road Side Unit}
	\acro{VANET}{Vehicular Ad Hoc Network}
	\acro{V2V}{Vehicle to Vehicle}
	\acro{V2I}{Vehicle to Infrastructure}
	\acro{VTL}{Virtual Traffic Lights}
	\acro{ETSI}{European Telecommunications Standards Institute}
	\acro{UMTS}{Universal Mobile Telecommunications System}
	\acro{LTE}{Long Term Evolution}
	\acro{eNodeB}{evolved Node B}
	\acro{LAN}{Local Area Network}
	\acro{CH}{Cluster Head}
	\acro{QoS}{Quality of Service}
	\acro{RSS}{Received Signal Strength}
	\acro{UMTS}{Universal Mobile Telecommunication System}
	\acro{FCD}{Floating Car Data}
	\acro{CDF}{Cumulative Distribution Function}
	\acro{SUMO}{Simulation of Urban MObility}
	\acro{CI}{Confidence Interval}
\end{acronym}

\begin{abstract}
Vehicular Networks enable a vast number of innovative applications, which rely on the efficient exchange of information between vehicles. However, efficient and reliable data dissemination is a particularly challenging task in the context of vehicular networks due to the underlying properties of these networks, limited availability of network infrastructure and variable penetration rates for distinct communication technologies. This paper presents a novel system and mechanism for information dissemination based on virtual infrastructure selection in combination with multiple communication technologies. The system has been evaluated using a  simulation  framework, involving network simulation in conjugation with realistic vehicular mobility traces. 
The presented simulation results show the feasibility of the proposed mechanism to achieve maximum message penetration with reduced overhead.  
Compared with a cellular-based only solution, our mechanism shows that the judicious vehicle selection can lead to improved network utilization through the offload of traffic to the short-range communication network.
\end{abstract}

%
\IEEEpeerreviewmaketitle

\section{Introduction} \label{sec:intro}

Vehicular Networks enable a vast number of innovative applications, namely safety, traffic efficiency and information/entertainment applications.
These applications rely on the exchange of information between nodes and can greatly benefit from information generated far away (e.g. to warn drivers of accidents and road works ahead).
Communication in Vehicular Networks comprises of \ac{V2V} and \ac{V2I} communications based on short-range wireless \ac{LAN} or infrastructure-based networks (e.g. \ac{LTE}).
Depending on the application characteristics and requirements, information can be propagated locally between vehicles (possibly via multiple hops) and/or 
can be supported by infrastructure networks or a central entity for prorogation of messages over larger distances.

Efficient and reliable data dissemination is a challenging task, which is specially evident  in the context of Vehicular Networks.First, the underlying properties of \acp{VANET}, such as the variability of the network topology, high speed of the vehicles, network partitioning into clusters of vehicles,constrained vehicle mobility and uneven network density, create additional challenges for timely data dissemination in large scales. Second, limited static infrastructure availability can impair \ac{V2I} communications. For instance for \acp{VANET}, it is foreseen that dedicated infrastructure will be limited or even non-existing in some regions during the initial deployment phases. Third, data dissemination based on a single technology paradigm can limit the solution optimality since \ac{V2V} and \ac{V2I} technologies have different advantages and drawbacks (\cite{lteVehNet}). For example, cellular networks can have higher coverage but will also deliver lower performance in terms of latency compared to short-range communication networks. Also the information to be disseminated in a geographic region can be intermittent and relatively smaller in size. This may induce overhead on the cellular network leading to improper channel utilization. In previous works these different technologies have been mainly presented as alternative communication means. Four, the dissemination process should also take into consideration the variable penetration rates for the several communication technologies.

To address the above mentioned challenges, in this work we propose a novel system and mechanism for information dissemination in multi-technology vehicular networks. Our main goal is to increase the penetration of information in a geographic region efficiently and reliably. In this paper, we argue that this can be achieved by the assistance of virtual infrastructure, i.e. mobile (either stationary or moving) infrastructure nodes. On one hand selecting appropriate vehicles as virtual infrastructure can alleviate the requirements for fixed infrastructure . On the other hand, by combining multiple communication technologies, their advantages in terms of characteristics and performance could be combined while still considering variable penetration rates.  To optimize the election of vehicles as virtual infrastructure an optimal dissimilarity relation defined among the vehicles under constraints, such as vehicle mobility, network load, application requirements. The proposed algorithm is evaluated under various criterion to realistically provide evidence that such a method can improve the message penetration in a given geographic region with sparse infrastructure while keeping communication overheads at minimum.

The remainder of the paper is organized as follows. In Section~\ref{sec:related} we provide the relevant related work in the broad area of information dissemination in vehicular networks. In section~\ref{sec:algorithm}, we present in detail the proposed system with architecture and method for electing virtual infrastructure to disseminate information efficiently and adaptively in a multi-technology vehicular network. Section \ref{sec:evaluation} explains the simulation environment and the selected evaluation metrics. The evaluation of the proposed algorithm are discussed in section \ref{sec:results}.  Finally, section~\ref{sec:conclusions} gives the main conclusions and future research.

\section{Related Work} \label{sec:related}

In \acp{VANET} information dissemination mechanisms allow drivers to be aware in real-time of their surroundings. Extensive research has been conducted in the broad area of information dissemination, with a main objective of transferring data in a reliable manner between nodes participating in tje communication network while meeting certain design objectives. Design objectives may include low delay, high reliability, low overhead, among others. 

The vast majority of previously proposed methods focus on single-technology data dissemination. In many solutions improvement in information dissemination is achieved through deployment of infrastructure nodes at preferential locations (e.g. intersections, busy road segments). 
The problem of selecting the appropriate set of locations for static infrastructure nodes has been studied in several previous works \cite{Lochert2009b}.
However, several factors, including cost, complexity, existing systems, and lack of cooperation between government and private sectors, have impeded the deployment of RSUs~\cite{carRSUTonguz}.
Additionally,  these static placements lack the flexibility and may not provide the desired dissemination coverage due to the dynamic nature of vehicular networks.

Few works in the literature propose to utilize vehicles as temporary \acp{RSU}.C\^amara et al. \cite{camara} present the virtual RSU (vRSU) concept where nodes receive and cache messages from other vRSUs or access points which are propagated in areas with no coverage from conventional \acp{RSU}. 
Eckhoff et al. \cite{parkedCars}  introduce the concept of using parked cars as relay nodes in vehicular networks in especially challenging propagation conditions (e.g. urban intersections). 
Tonguz et. al \cite{carRSUTonguz} propose a distributed algorithm for selecting vehicles as temporary \ac{RSU}s, which stop for a short time interval for rebroadcasting messages, based on the direction (towards accident) and location within the region of interest (boundary). 

Another common technique for  data dissemination, relies on the creation of self-organized and dynamic clusters enabled by short-range communications \cite{chai}\cite{Cherif}\cite{TFcluster}.
These clusters are constituted by a number of  members and cluster head(s), which control and/or execute the dissemination process
and perform inter-cluster communication.
Vehicles create and maintain clusters depending on a number of metrics (e.g. link quality, vehicle direction or speed) by periodically exchanging status information. However, such periodic exchange of messages which can create considerable additional data traffic in the network. Chai et al. \cite{chai} propose a method for \textit{cluster head selection} based on node degree, available resources and vehicle mobility, 
and a scheme for \textit{cluster switching} based on a utility function based on \ac{QoS} requirements. 
However, in these proposals cluster heads are selected independently, which impacts several network parameters (e.g. efficient resource usage).

More recently research in vehicular networking has focused on data dissemination in heterogeneous networks.
In general, in hybrid networks nodes use \acp{VANET} for \ac{V2V} communication (e.g. ITS G5, 802.11p/WAVE \cite{dsrc09}) 
and cellular networks for \ac{V2I} communications. Many works in this area also make use of cluster-based mechanisms
where vehicles are grouped into clusters according to selected parameters.
Majority of proposals have focused on \textit{distributed} cluster formation and \ac{CH} or gateway election (e.g. \cite{Taleb}\cite{qosGateway}\cite{sivaraj})
Benslimane et al. \cite{Taleb} delineate a  scheme for dynamic clustering of vehicles based on the direction of vehicles’ movement,
 \ac{UMTS} \ac{RSS}, and inter-vehicular distance.
In \cite{qosGateway} Zhioua et al. provided a multi-metric QoS-balancing gateway selection mechanism  in hybrid vehicular networks depending on the data to be transmitted.
In \cite{tung} Tung et al. propose a collision avoidance service that relies on Wi-Fi for local information dissemination and cluster creation,
which are used by \ac{LTE} for inter-cluster communication.  Li et al. introduce in \cite{coalition} a cooperative protocol for efficient data dissemination based on coalition game theory.
 
\textit{Centralized} cluster management (i.e. creation and management) has also been proposed as an efficient method for information dissemination.
In \cite{LTE4V2X} Remy et al. put forward a new paradigm for vehicular network organization based on \textit{centralized} cluster management using multiple technologies. 
This approach makes use of \ac{FCD} received at an \ac{eNodeB} to setup and maintain clusters but still relies on \ac{VANET} for advertisement of cluster information. 
Simulation results show performance improvements in terms of lower overhead, improved goodput and smaller packet loss when comparing with a decentralized approach.
Thus, this centralized scheme can lead to improved cluster formation and management since there is additional information for decision making. However, in most of the proposed algorithm do not address how information penetration can be improved in vehicular networks, which is one of the key criteria for the success of vehicular networks. To achieve the required message penetration while considering the application and vehicular network requirements is the focus of this work.



\section{Multi-technology information dissemination} \label{sec:algorithm}
\begin{figure*}[t!]
\centering
\includegraphics[trim=0.7cm 6.5cm 0.7cm 3cm,clip=true,width=0.85\textwidth]{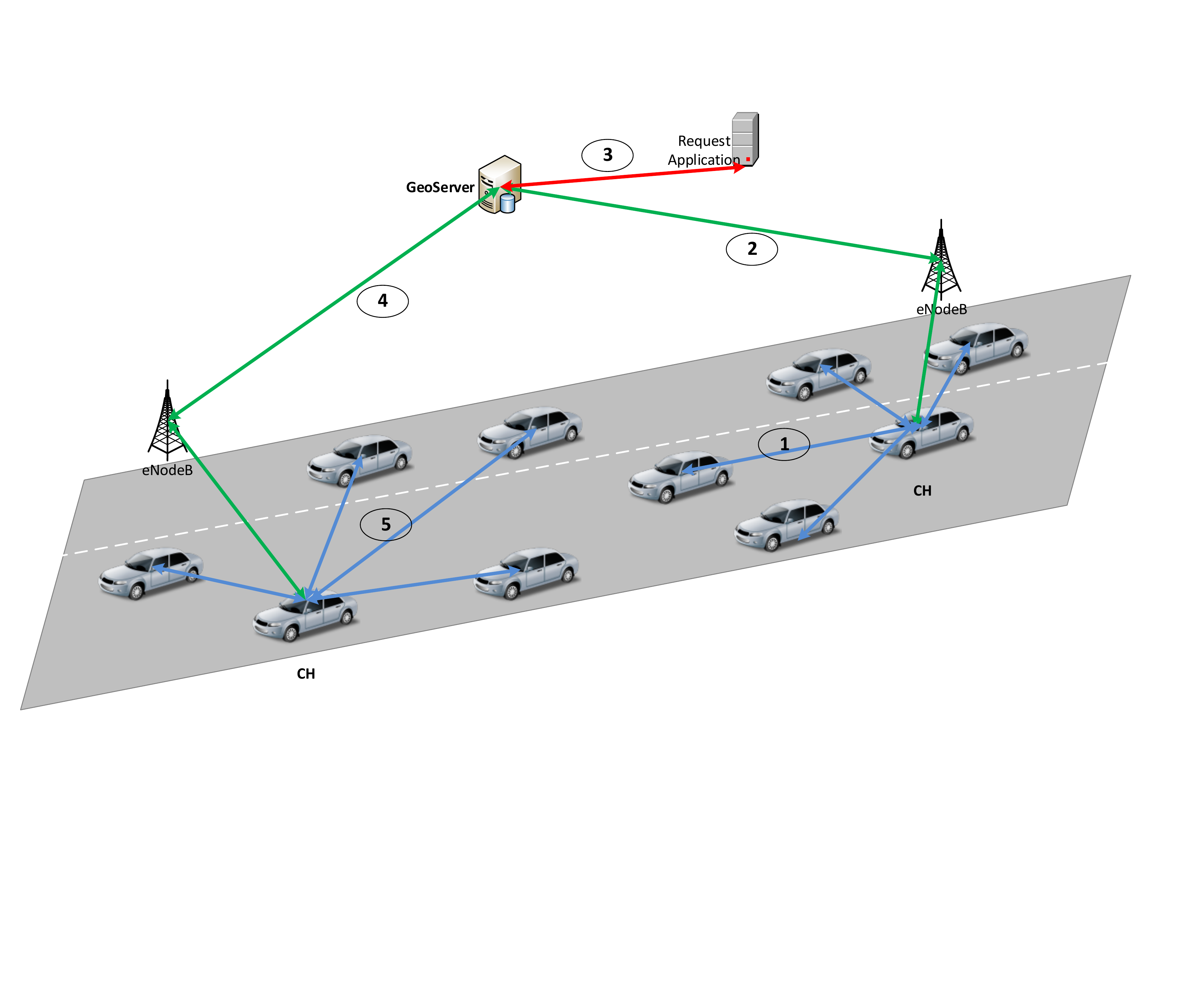}
\centering
\caption{Multi-technology information dissemination system, which comprises three main phases (a) data collection, (b) virtual infrastructure selection and (c) data dissemination strategy execution. In more detail, the system comprises the following phases  (1) Periodic broadcast of \acp{CAM} allows collecting information on the dynamic neighborhood relation. 
(2) Communication to geoserver of aggregated neighbor tables; (3) Data dissemination request and virtual infrastructure selection at geoserver; 
(4) information dissemination execution and (5) local data dissemination by virtual infrastructure.}		
\label{fig1}
\end{figure*}	


In this paper, we propose a system and mechanism to provide maximum information penetration in a defined geographical area, which is specially useful in scenarios with sparse infrastructure deployment. The multi-technology information dissemination system relies on the collection of neighborhood information collected at vehicles  to determine the best data dissemination strategy at a centralized location. The system selects mobile (stationary or moving) infrastructure nodes (vehicles) in a multi-technology vehicular environment, 
including short-range communication networks (e.g. ITS-G5) and long-range communication networks (e.g. cellular).
The selection of vehicles to act as virtual infrastructure is based on application requirements, the characteristics of vehicular environment, network load, among other constraints.
The proposed greedy approach ensures minimal computational to maximize the election efficiency.

\subsection{Architecture}

Figure \ref{fig1} outlines the general architecture of the proposed system that comprises of three main entities: a central entity (i.e. Geoserver), 
Infrastructure Units (RSU's and eNodeB) and Vehicles. In the proposed approach vehicles are assumed to be equipped with a positioning system, 
short-range and/or long-range communication capabilities. In addition, vehicles exchange information that enables building neighbor tables.
 In terms of execution, the multi-technology information dissemination system comprises three main phases: 
 (a) data collection, (b) virtual infrastructure selection and (c) data dissemination strategy execution. 
 In the following, we present in more detail each of these phases.

\paragraph{\underline{Data Collection}} 
Vehicles broadcast periodically broadcast single-hop \acfp{CAM} that contains static (e.g. vehicle dimensions) and dynamic (e.g. position, speed) 
vehicle information. By receiving \acp{CAM}, vehicles become aware neighbor stations as well as their positions, movement, basic attributes and basic sensor information.
 This enables the construction of neighbor tables at each vehicle containing at least the received message in conjugation with a timestamp.
To further enable the capabilities of the algorithm, we argue that \acp{CAM} should be enriched with information such as 
available communication technologies, neighbor tables, etc.
For instance, including neighbor tables in these periodic beacons could provide an extended view of the neighborhood.
As decision making is done at a central location, 
these aggregated neighbor tables are transmitted to the Geoserver to serve as basis for the virtual infrastructure election procedures.


\paragraph{\underline{Virtual Infrastructure Selection}}
The information collected allows the geoserver to have a bird-eye view of static and dynamic network characteristics. 
The geoserver may also receive additional information from other data sources (e.g. coverage information from network operators). 
In the proposed method, the selection of the  virtual infrastructure node is made at the central entity, i.e. Geoserver.
 The Geoserver is connected to the infrastructure units and has an interface for receiving requests from service providers.
  Request for disseminating information should contain at least the dissemination area and, optionally, the data to be transferred and additional constraints. 
 Whenever, the geoserver receives the request from the service provider, the virtual infrastructure selection process starts.
 Based on input requirements and the data collected in step 1,
  the Geoserver analyzes all potential vehicle \ac{ITS} stations that can act as virtual infrastructure 
  and iteratively selects the nodes that maximize the message penetration in a given geographic area. 
 In the selection process, the Geoserver considers static and dynamic constraints, 
 namely service requirements, network constraints and vehicular network constraints prior to making the selection decision.
 More details on this mechanism are provided in Section~\ref{sec:alg}.


\paragraph{\underline{Data Dissemination Strategy Execution}}
The decisions made at the geoserver are propagated to selected vehicles that perform local action execution. 
The selected vehicles can also instruct other nodes to further propagate the information among the peers.
The generic method also considers the adaptation of the geoserver instructions at vehicles if the local conditions have evolved.

To conclude, the main functions of the Geoserver can be summarized as follows:
\begin{itemize}
\item receive, store and process information coming from vehicles and other data sources;
\item to periodically estimate vehicle nodes and infrastructure nodes coverage area whenever this information is not provided by a third party. 
\item determine the best strategy for multi-technology information dissemination in a geographical area taking into consideration the service requirements and other static and dynamic information. This step be executed either periodically or whenever a request for information dissemination is received. 
\end{itemize}

\subsection{Algorithm} \label{sec:alg}

The proposed algorithm is formulated as a constrained version of the maximum/minimum optimization problem.
The key optimization objective is to  maximize the coverage area (e.g. number of receiving nodes) with minimum set of nodes while considering several constraints.
Nodes are selected from the set of available nodes based on the \textit{dissimilarity index}. 
The dissimilarity relation is an index where the less similar areas a vehicle covers, the larger are the relation index values. 
The proposed dissimilarity relation between vehicles in a given geographic region has an influence on the selection procedure of the virtual infrastructure at the geoserver. 
During the selection procedure, all the constraints (application requirements, network load, vehicle mobility, etc.) that originate from different entities should be considered. 
To compute the dissimilarity relations , the geographical area is assumed to be divided into a number of (adaptive) sub-areas.

\subsubsection{Computation of the Dissimilarity Index}

As detailed the previous Section, vehicles ($V_i$) update their neighbor tables (NT) and communicate them to the Geoserver.
Once the information from vehicles in a geographic region has been aggregated at the Geoserver, it builds a dissimilarity relation index ($D_i$) between the potential vehicle ITS stations that can act as virtual infrastructure. The basic concept for the dissimilarity relation index is that the less similar areas a vehicle covers, the larger the relation index value. This allows selecting a minimal set of vehicles to act as virtual infrastructure while maximizing the coverage area.  To select the first vehicle station in the set, a similar procedure is applied, 
 but instead each vehicle computes it's self-correlation index , which was denominated Zone Index ($Z_i$). The zone index is defined as the number of regions a vehicle acting as virtual infrastructure covers in either a single iteration or multiple iterations (limited by hop count). Along with $Z_i$, the selection procedure is subjected to other constraints like application requirements, network load, vehicle mobility, etc. 
 The vehicle with the highest Zone Index is then selected as the
 first Virtual Infrastructure Node.

\begin{figure}[t!]
 		\centering
		\includegraphics[width=0.5\textwidth]{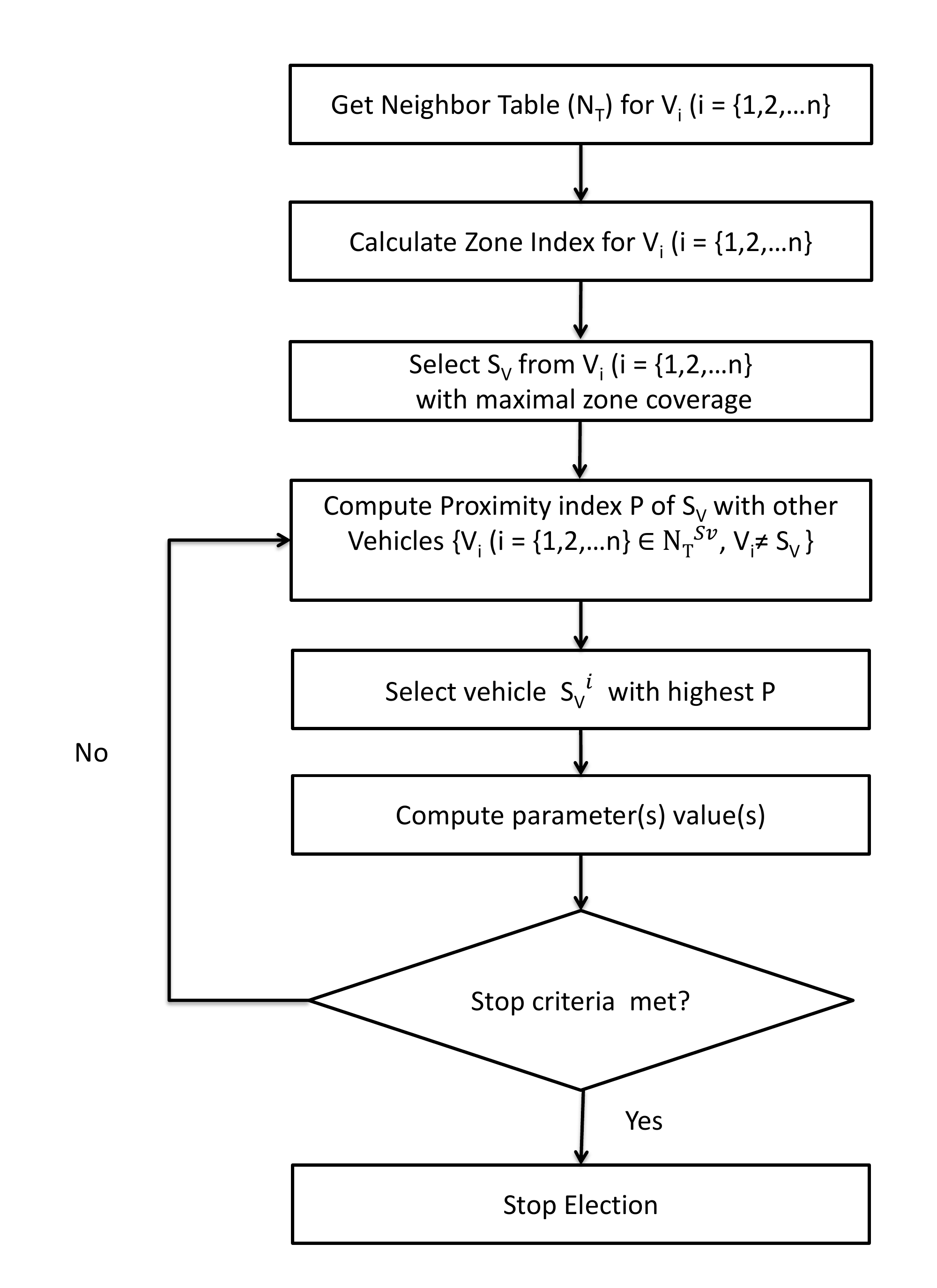}
		\centering
		\caption{Flow Chart for Virtual Infrastructure Selection}		
		\label{fig_Infra}
		\end{figure}

The procedure for selecting vehicles to become virtual infrastructure (relay, forwarder) nodes using the dissimilarity index is detailed further in Figure \ref{fig_Infra}. Once the first vehicle is selected, the node with the highest dissimilarity index is selected as the next best candidate as virtual infrastructure. Dissimilarity index is determined by the number of non-overlapping zones, based on criteria set, either in a single or multi-hop. Upon election, the procedure can be further iterated or stopped based on any criteria derived from the constraints. In the following we provide some examples for the stopping criterion:
\begin{itemize}
	\item information depreciation: validity time of the information set by the application;
	\item number of virtual infrastructure nodes dependent of the resource availability and current/instantaneous network conditions.
\end{itemize}

To conclude, the proposed method determines the best set of virtual infrastructure nodes to maximize data dissemination in a geographical area while minimizing resource consumption. Depending on request requirements (e.g. in terms of delay), the algorithm can adaptively 
select through two dimensions: i) number of nodes and ii) number of hops.



\section{Simulation-based Evalution of System Performance} \label{sec:evaluation}
\subsection{Methodology}\label{sec:meth}

\begin{figure*}[!t]
	\centering
		\includegraphics[width=0.75\textwidth]{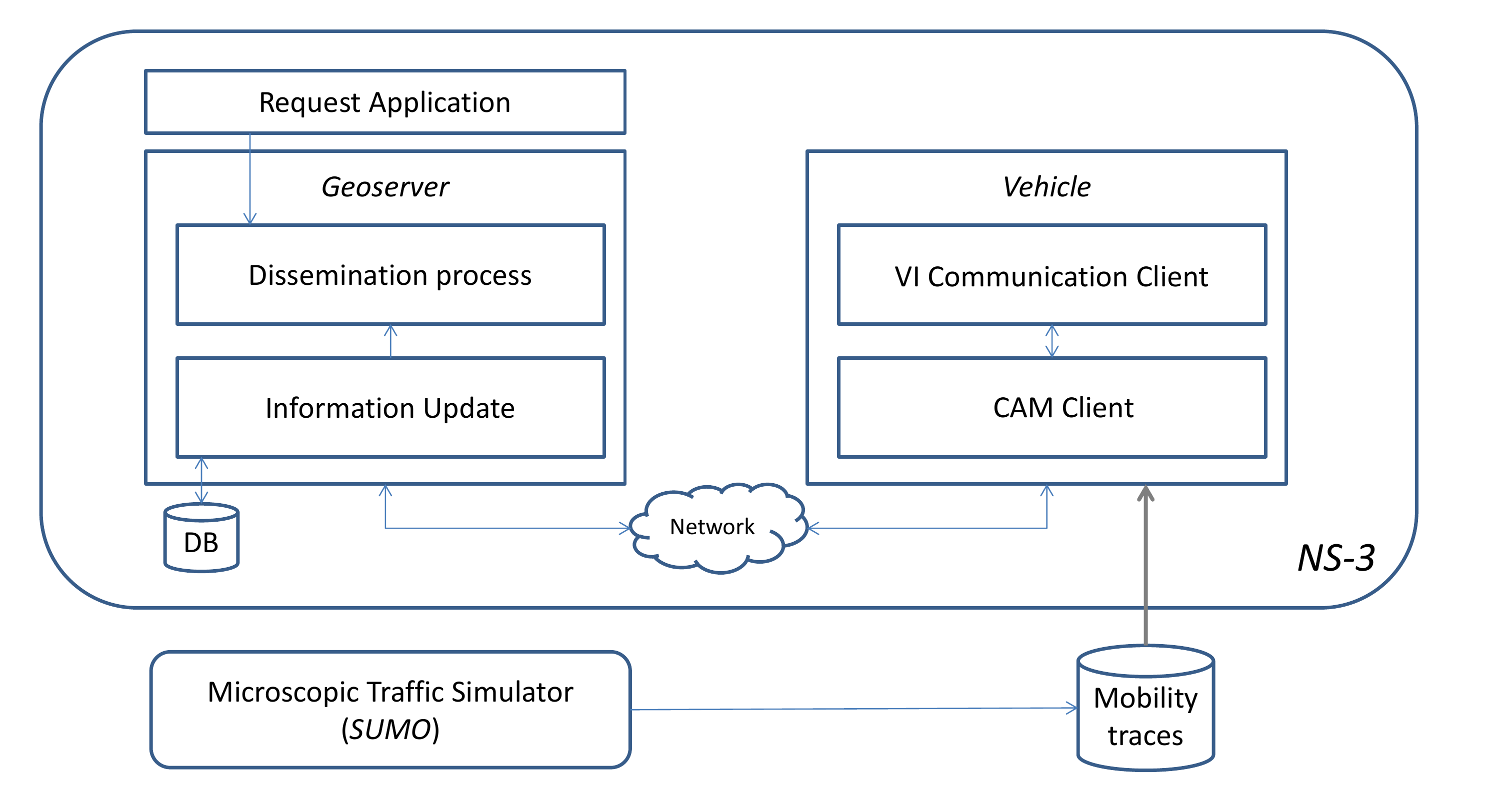}
	\caption{Simulation platform. The main component is the NS-3 discrete network simulator, which mimics the behavior and the interaction between \textit{Vehicles} and the \textit{Geoserver} (central entity for decision making). Realistic vehicular mobility traces (generated using the \acs{SUMO} microscopic traffic simulator) are fed as input to NS-3.  }
	\label{fig:simulation}
\end{figure*}

The proposed data dissemination algorithm is evaluated by means of simulation. 
With this intent the system has been implemented in the network simulator NS-3~\cite{ns3Site}\cite{henderson2008network} including  a comprehensive set of functions. 
NS-3 is an open source and validated discrete-event network simulator, which was primarily designed for Internet systems.
This network simulation has been selected due to its complete set of features and the superior performance as demonstrated in the study by Weingartner et al.~\cite{NSperf}. 
The latest version of this network simulator (NS-3.19) provides a complete set of models for assessing heterogeneous vehicular networks, 
 including models for \ac{LTE} communication networks and short-range vehicular communication networks (i.e.  802.11 p). 


Fig.~\ref{fig:simulation} depicts a simplified overview of the simulation platform.
The NS-3 network simulator has been extended with two new modules, Network information update module and Information dissemination module,
that are necessary for the algorithm implementation and evaluation. The main functions of each of these modules are presented in the following in more detail:

\begin{itemize}
\item \textit{Network information update module}: responsible for monitoring the evolving network properties and providing the Geoserver with updated network information.
The functionality of this module is distributed between vehicles and the geoserver. 
Vehicles implement a simplified version of the \ac{CAM} standard~\cite{etsistd}.
 Each vehicle periodically broadcasts single-hop \acp{CAM} messages containing static and dynamic information (e.g. position, speed).
By receiving periodic \acp{CAM}, every vehicle is aware of other stations in its neighborhood, which allows constructing neighbor tables at the receiving end.
On the server side, this information is received and processed to be used as input for the decision making and execution module.


\item \textit{Information dissemination module} : responsible for decision making and strategy execution. 
This module has the following three main functions: i) to listen and manage dissemination requests, ii) to determine the most appropriate data dissemination strategy
by electing vehicles as virtual infrastructure and iii) to execute the strategy by disseminating the information into and in the dissemination area.
The dissemination of information by initiated by the geoserver (to several nodes) and further executed by these nodes. 

\end{itemize}
To implement the above mentioned modules it is required to establish the complete communication cycle from data collection, 
aggregation to dissemination strategy determination and enforcement at the Geoserver. The messages exchanged between the different entities are the following (Fig.~\ref{fig1}):
\begin{itemize} 
	\item \textit{vehicle $\rightarrow$ vehicle}: information exchange between vehicles, including \ac{CAM} and local dissemination of request data (1 \& 5); 
	\item \textit{vehicle $\rightarrow$ Geoserver}: Neighbor table updates (2); 
	\item \textit{application $\rightarrow$ Geoserver}: application request that triggers the information dissemination process (3); 
	\item \textit{Geoserver $\rightarrow$ vehicles}:  virtual infrastructure notification (4) and enforcement of the solution by selected vehicles (5).
\end{itemize}

Realistic vehicular mobility traces is used as input for the network simulation platform. Mobility traces can be generated by a traffic simulator (e.g. SUMO) or  gathered during real world activities (e.g. making use of \ac{GPS}). A popular choice for generating offline mobility traces is \ac{SUMO}, an open source, highly portable, microscopic and continuous road traffic simulation package designed to handle large road networks~\cite{sumo}. Alternatively, bidirectionally coupled network and traffic simulators will be considered in future works.

\subsection{Metrics} \label{sec:metrics}
To evaluate the proposed algorithm following metrics have been defined:
\begin{itemize}
	\item \textbf{Covered Area ($\%$)}: the ratio of the number vehicles that have received the dissemination message to the total number of vehicles in the specified dissemination area. This metric describes the capabilities of the algorithm to maximize data dissemination in a given geographical area.  
	\item \textbf{Virtual Infrastructure Usage}: the number of selected virtual infrastructure nodes.
	 This metric allows understanding the ability of the algorithm to minimize resource consumption.
	\item \textbf{Overhead ($\%$)}: the total traffic generated as a result of the dissemination process, including request packets. 
	This metric will be presented combined for both  communication technologies (i.e. ITS G5 and \ac{LTE}) and allows understanding the overhead created by the process. 
	 	\item \textbf{Delay (ms)}: the time elapsed between the message transmission by the geoserver and the first message reception for a given vehicle. This metric allows understanding the temporal performance of the algorithm.
\end{itemize}

To evaluate the proposed algorithm in a realistic scenario, we consider in the simulation how distinct:
\begin{itemize}
	\item \textit{vehicle distribution patterns} influence the system performance for a given vehicle density.	
	\item \textit{vehicle densities} impair the mechanism ability to efficiently disseminate information in a geographical area.
	\item \textit{transmission power levels} affect neighborhood awareness and consequently the performance of the data dissemination algorithm.
\end{itemize}

We compare the performance delivered by our mechanism  against an All-LTE solution where there exists a permanent  bi-directional connection between \ac{eNodeB} and vehicles. 
Therefore, when an dissemination request is triggered, Geoserver sends separate messages to all vehicle in the dissemination area through \ac{LTE}.
This will result in optimistic results for the All-LTE solution since it assumes that all vehicles are equipped with at least \ac{LTE} technology.


\section{Results \& Discussion} \label{sec:results}

This section presents and discusses the evaluation of the system performance under varying controllable conditions,
 i.e. varying the maximum allowed number of virtual infrastructure nodes and varying the maximum transmission power. 
 Thus, in this section the impact of controllable (e.g. tx power) and uncontrollable parameters (e.g. mobility) is assessed.

\begin{table}[t]
\centering
\caption{Main Simulation Parameters}
\begin{tabular}{c c c} 
\textbf{Type}				& \textbf{Parameter}		& \textbf{Value} \\ \hline \hline
Neighbor Information		& CAM Frequency				&  1~Hz \\
							& Neighbor Table Timeout	&  5~s \\ 
							& Server update Frequency 	& 1~Hz \\ \hline
Dissemination Request		& Frequency					& 1~Hz \\ 
							& Dissemination area		& 0.44~$km^2$ \\ \hline
Scenario					& Type						& Urban (Malaga, Spain) \\
							& Number of Vehicles		& 45 \\
							& Simulation Duration		& 180~s \\ 
							& Vehicle Speed				& 10-50~km/h \\ 
							& Vehicle Density			& 113 $veh/km^2$ \\ \hline
802.11p Network				& Bit Rate					& 6~Mbps \\
							& Bandwidth					& 10~Mhz \\
							& Frequency band			& 5.9~GHz \\
							& Maximum Tx Power			& [16, 21, 23]~dBm \\
							\hline
LTE Network					& eNodeB Tx Power			& 30~dBm \\
							& UE Tx Power				& 10~dBm \\
							& Propagation Model			& Friis \\
\end{tabular}
\label{table:simPar}
\end{table}

\subsection{Simulation Scenario}
To realistic perform \ac{VANET} simulation, we have resorted to publicly available vehicular mobility traces in NS-2 format,
which are used as input to the network simulation platform. 
These mobility traces have been generated by \ac{SUMO} using  realistic input data, including road network, vehicle routes, traffic lights, among others.
The simulated urban scenario is the downtown area of the city of Malaga. 
Vehicles move during 180~s in the road network of an area with approximately 600~m x 700~m. The maximum vehicle velocity is 50~km/h. 
Additional details on the simulation scenario can be found in~\cite{jamal}.


With respect to the data traffic generation we consider:
\begin{itemize}
	\item \textit{Neighbor information}: \acp{CAM} are generated with a frequency of 1~Hz. Neighbor table information is send to the central entity with a frequency of 1~Hz and neighbor table entries expire in vehicles after 5~s.
	\item \textit{Dissemination requests}: Dissemination requests are generated every second and the dissemination area matches the simulation scenario.
\end{itemize}

More details on the main simulation parameters are found in Table~\ref{table:simPar}.

\subsection{Results}

This section presents the results obtained 
making use of the simulation platform presented in Section~\ref{sec:meth}
and the set of metrics defined in Section~\ref{sec:metrics}. 
The system performance is assessed for each simulation under varying uncontrollable parameters (node mobility),
i.e. several dissemination request are  analyzed for each simulation. This allows us to determine the mean value 
and 95~$\%$ \acp{CI} of the selected metrics across several requests.
In addition, we study the system performance under varying controllable parameters, namely by varying the:
\begin{itemize}
	\item maximum allowed number of selected virtual infrastructure nodes ($k$). 
	\item transmit power
\end{itemize}

It should be noted that $k$ is the upper limit for the virtual infrastructure set size. However, 
 the algorithm will select a set with size smaller than $k$ if the stop criteria is met.

\subsubsection{Neighborhood Awareness Levels}
 \begin{figure}
        \centering
		\subfigure[\scriptsize Tx power = 16 dBm]{\label{fig:conGra16}\includegraphics[width=0.15\textwidth]{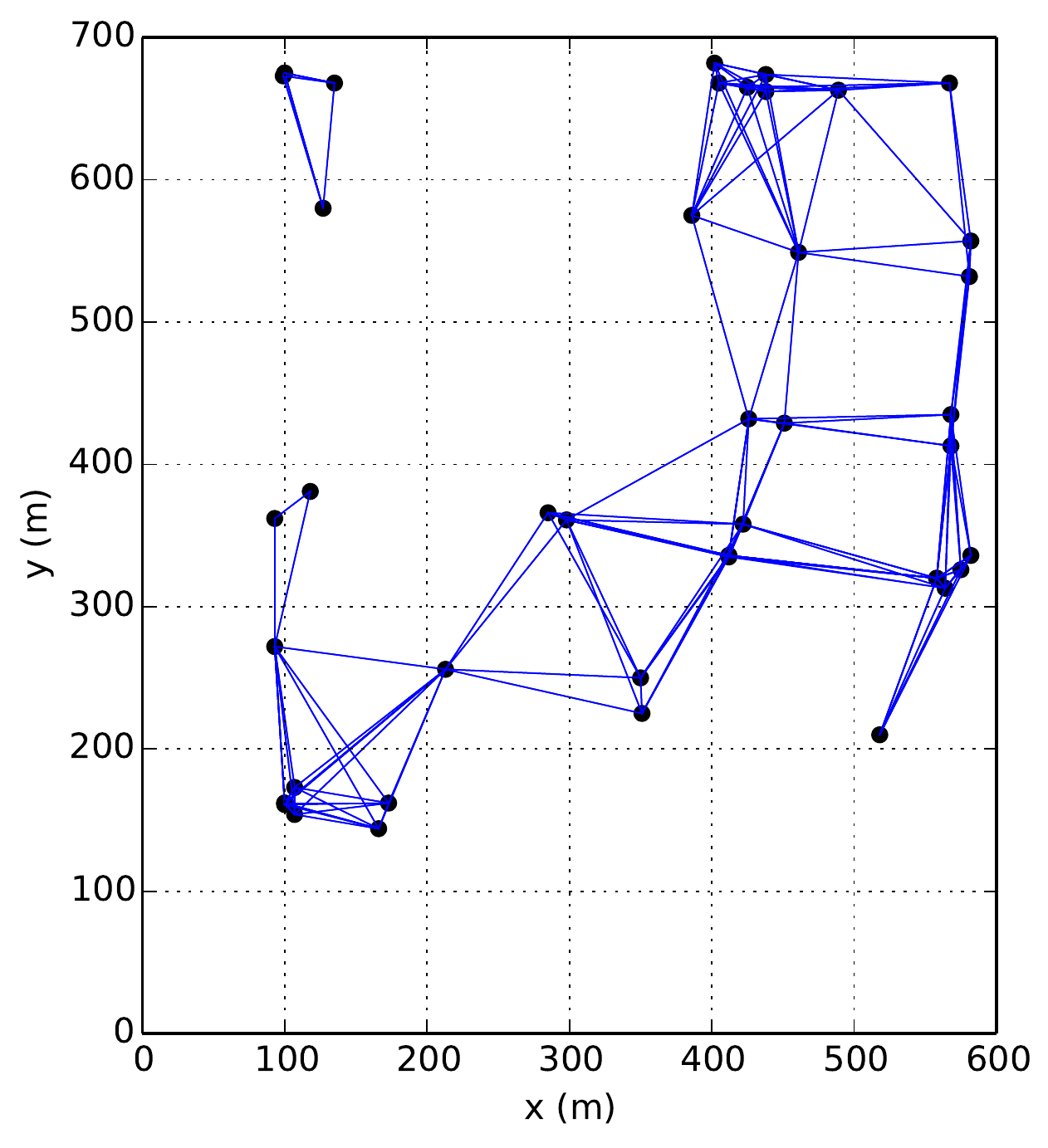}}
		\subfigure[\scriptsize Tx power = 21 dBm]{\label{fig:conGra21}\includegraphics[width=0.15\textwidth]{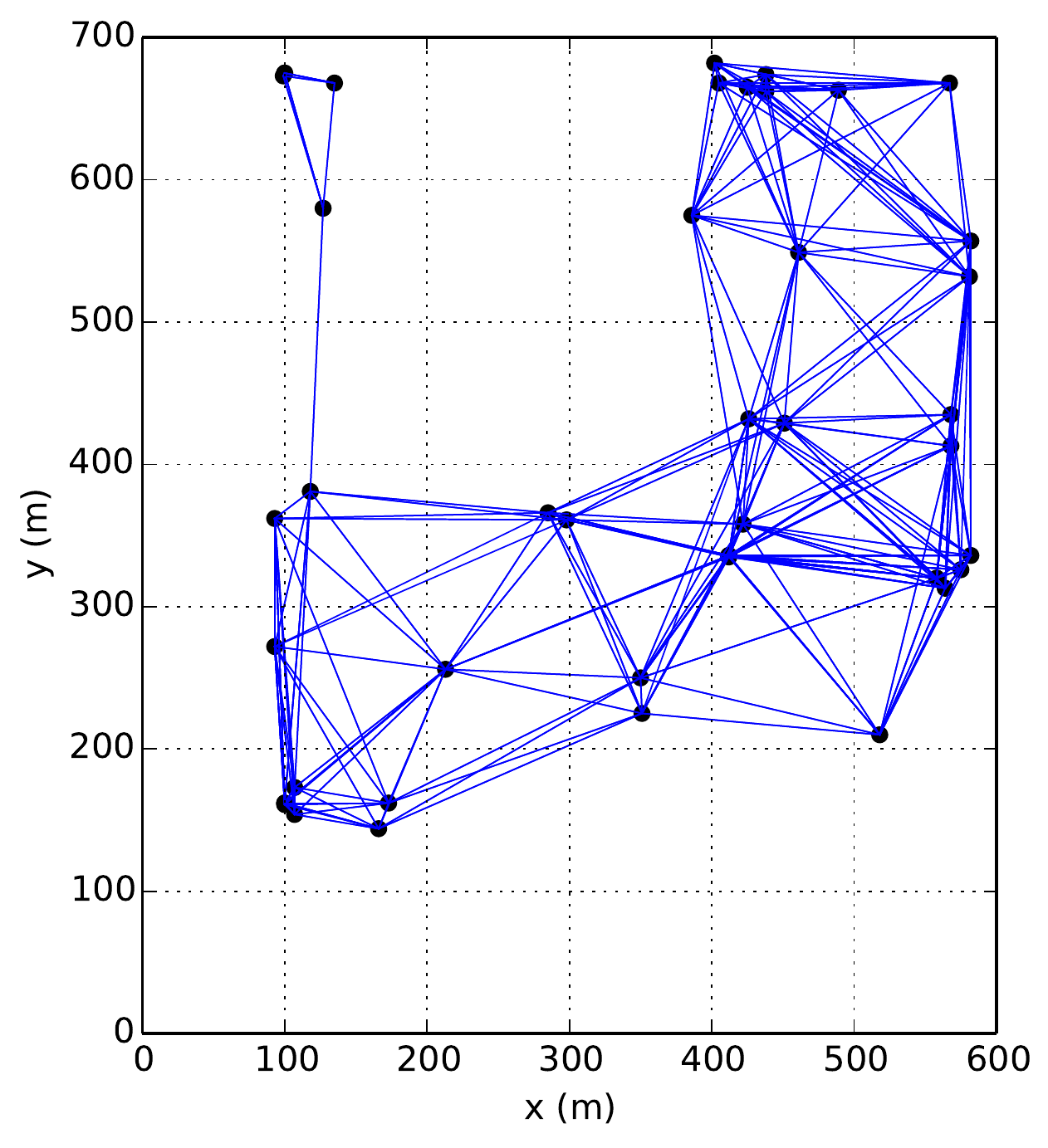}}
		\subfigure[\scriptsize Tx power = 23 dBm]{\label{fig:conGra21}\includegraphics[width=0.15\textwidth]{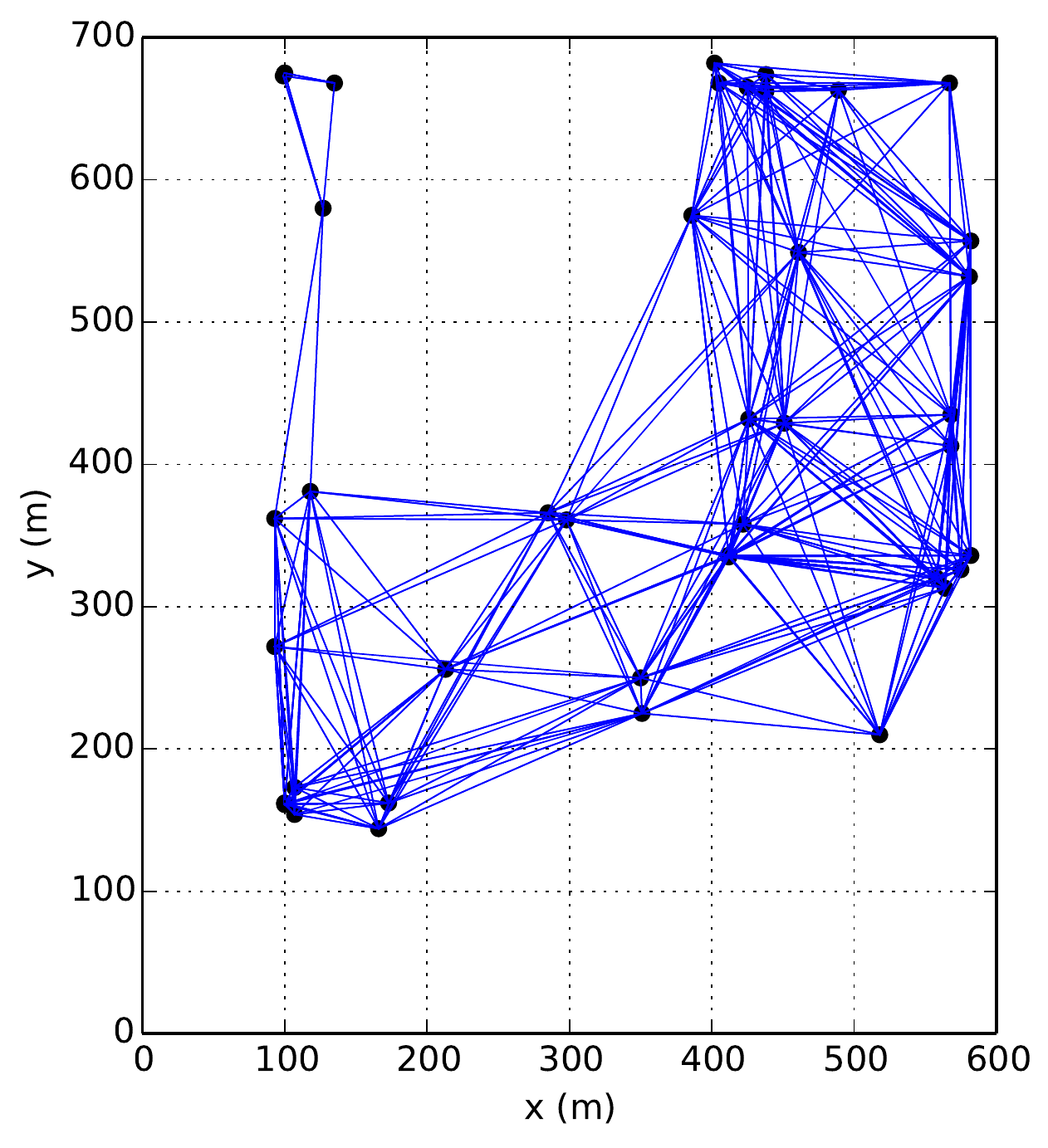}}
        \caption{Node connectivity graph examples for different maximum transmit power values (16, 21 and 23 dBm). 
        As expected, the node degree increases considerably for increasing transmission power due to the higher communication range.}
        \label{fig:INPUT}
\end{figure}

\begin{figure}[t]
        \centering
		\includegraphics[width=0.35\textwidth]{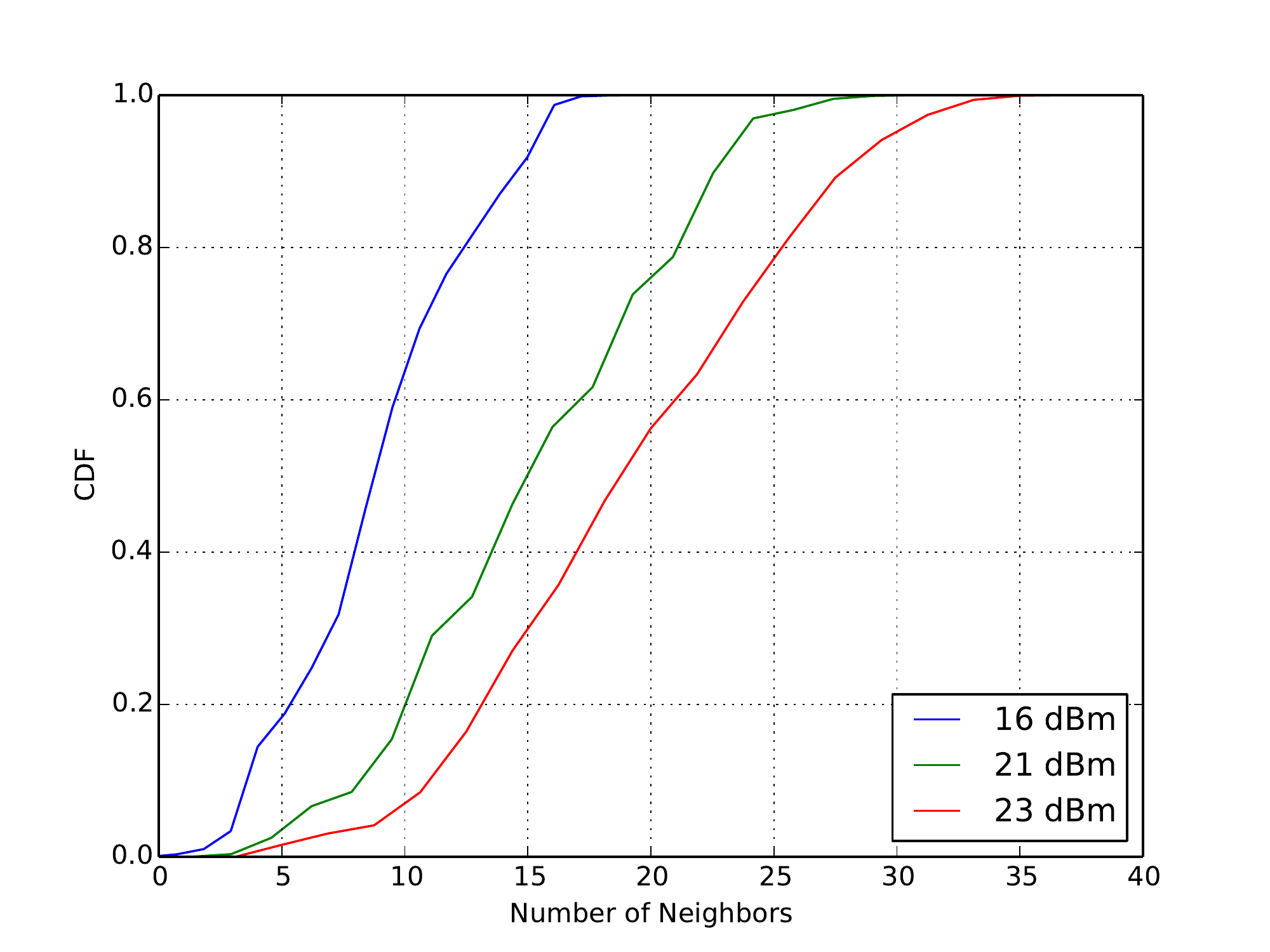}     
        \caption{Neighborhood awareness levels. Increasing the transmission power improves local neighborhood awareness levels.}
        \label{fig:NA}
\end{figure}

 To better understand the presented results we first provide a short overview of the input dataset.
As the proposed algorithm relies on information from surrounding vehicles, we  mainly focus on the neighbor awareness. 
Fig.~\ref{fig:INPUT} presents an example vehicle connectivity graph for different transmission powers values, ranging from 16 to 23~dBm, for a given timestamp $t$. 
As expected, increasing transmission powers extend the communication range and consequently neighborhood awareness levels at vehicles. 
More details on the relation between link quality (e.g.  in terms of \ac{PDR}) and neighborhood awareness levels can be found in~\cite{pedro}.
 Additionally, increasing transmission power improves connectivity and can reduce network partition.  
 Fig.~\ref{fig:NA} presents the \ac{CDF} of the number of neighbors for each node, which corresponds to the node degree in the network graph. 
We consider in the analysis the complete simulation duration and varying transmit powers. 
For the selected mobility trace, the median number of neighbors is approximately 10, 15 and 20 neighbors for 16, 21 and 23~dBm transmit power, respectively.
This shows that the network is fairly connected and that increasing transmit powers lead to higher node degree.
These results are inline with the work by Baumann et al~\cite{baumann2008generic}.
 
\subsubsection{Information Dissemination Results}

\begin{figure*}
        \centering
		\subfigure[\scriptsize Covered Area]{\label{fig:covVeh}\includegraphics[width=0.32\textwidth]{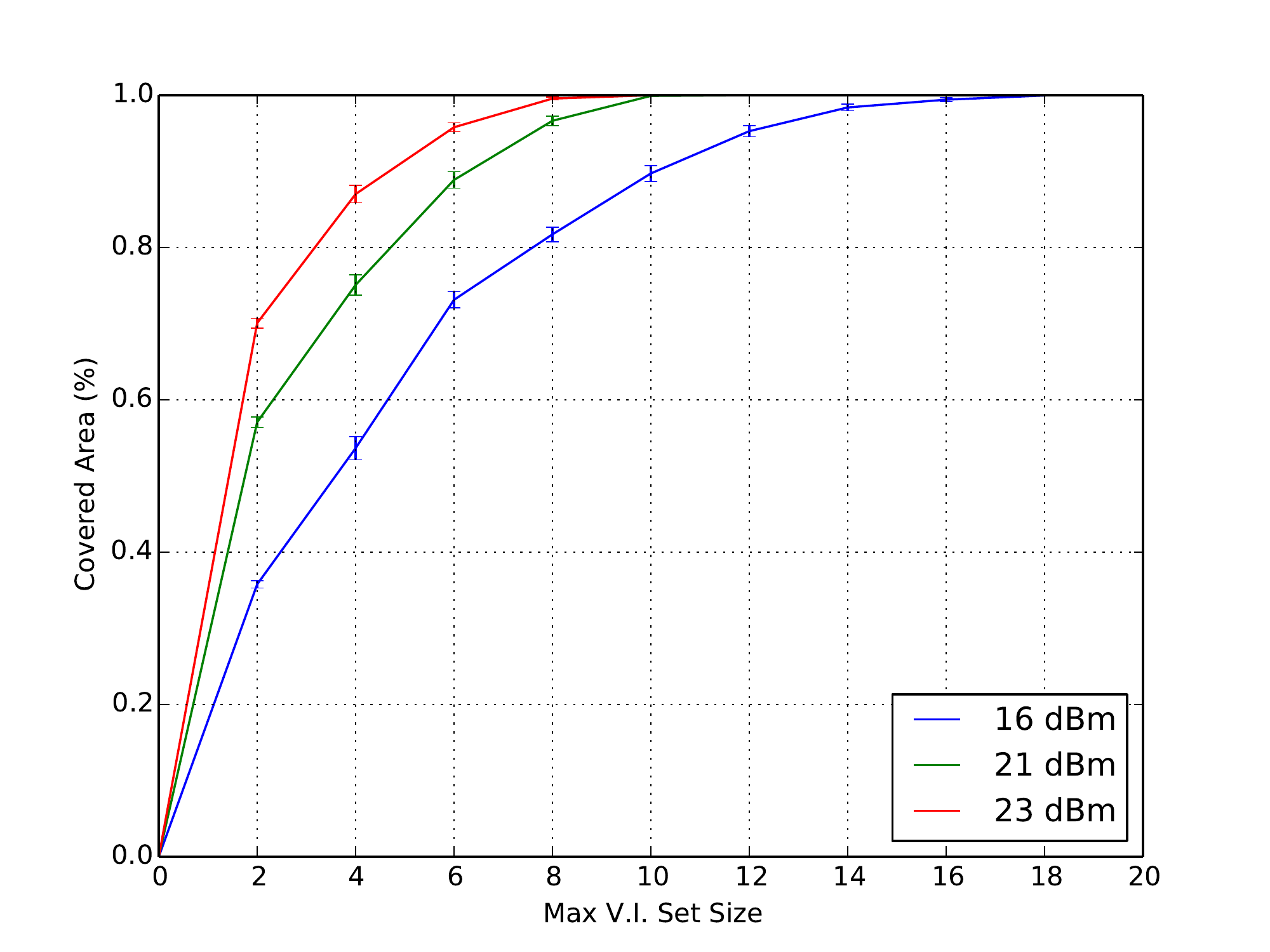}}
		\subfigure[\scriptsize Mean V.I. Set size]{\label{fig:chSize}\includegraphics[width=0.32\textwidth]{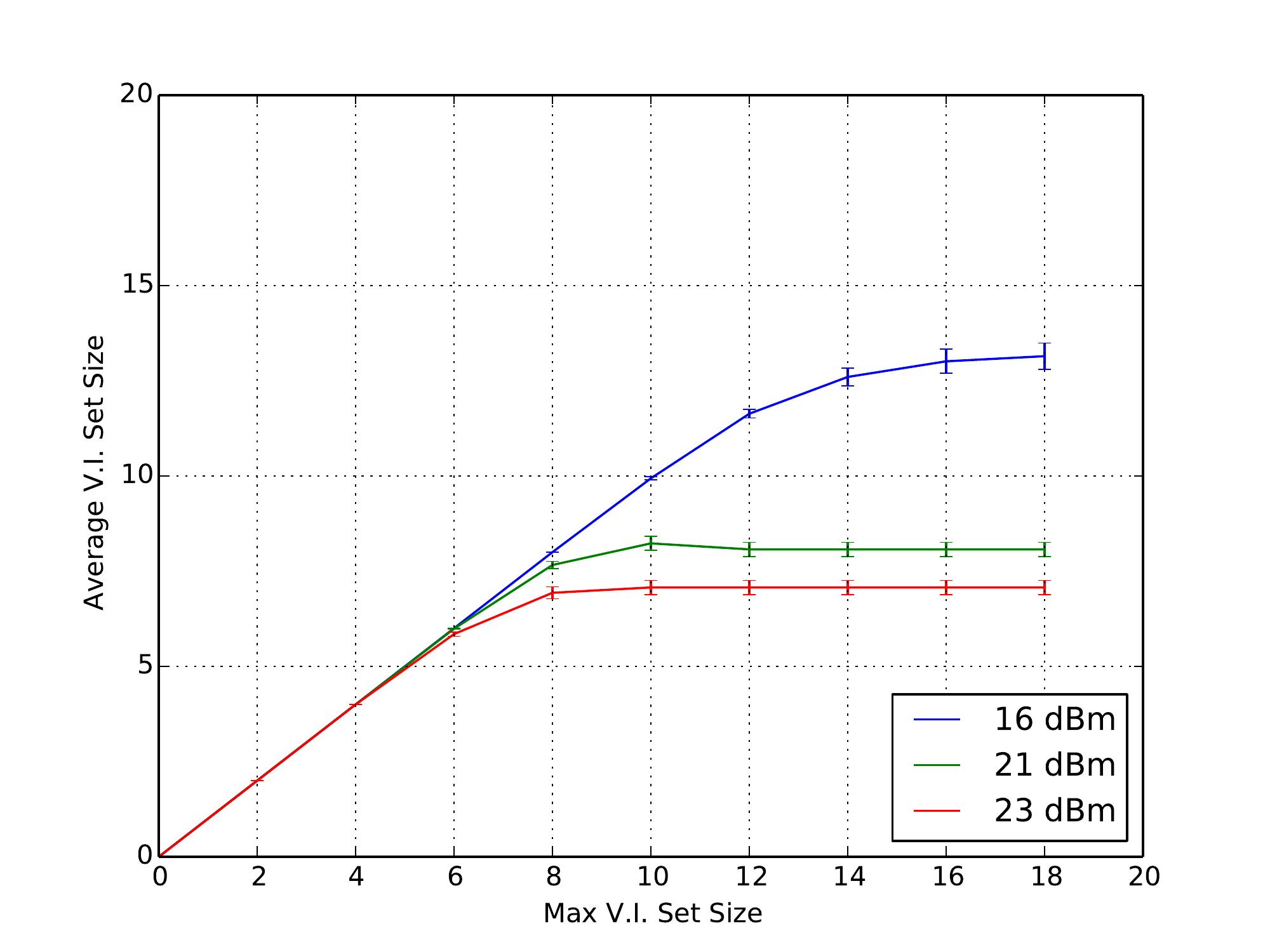}} 
		\subfigure[\scriptsize Overhead]{\label{fig:traffic}\includegraphics[width=0.32\textwidth]{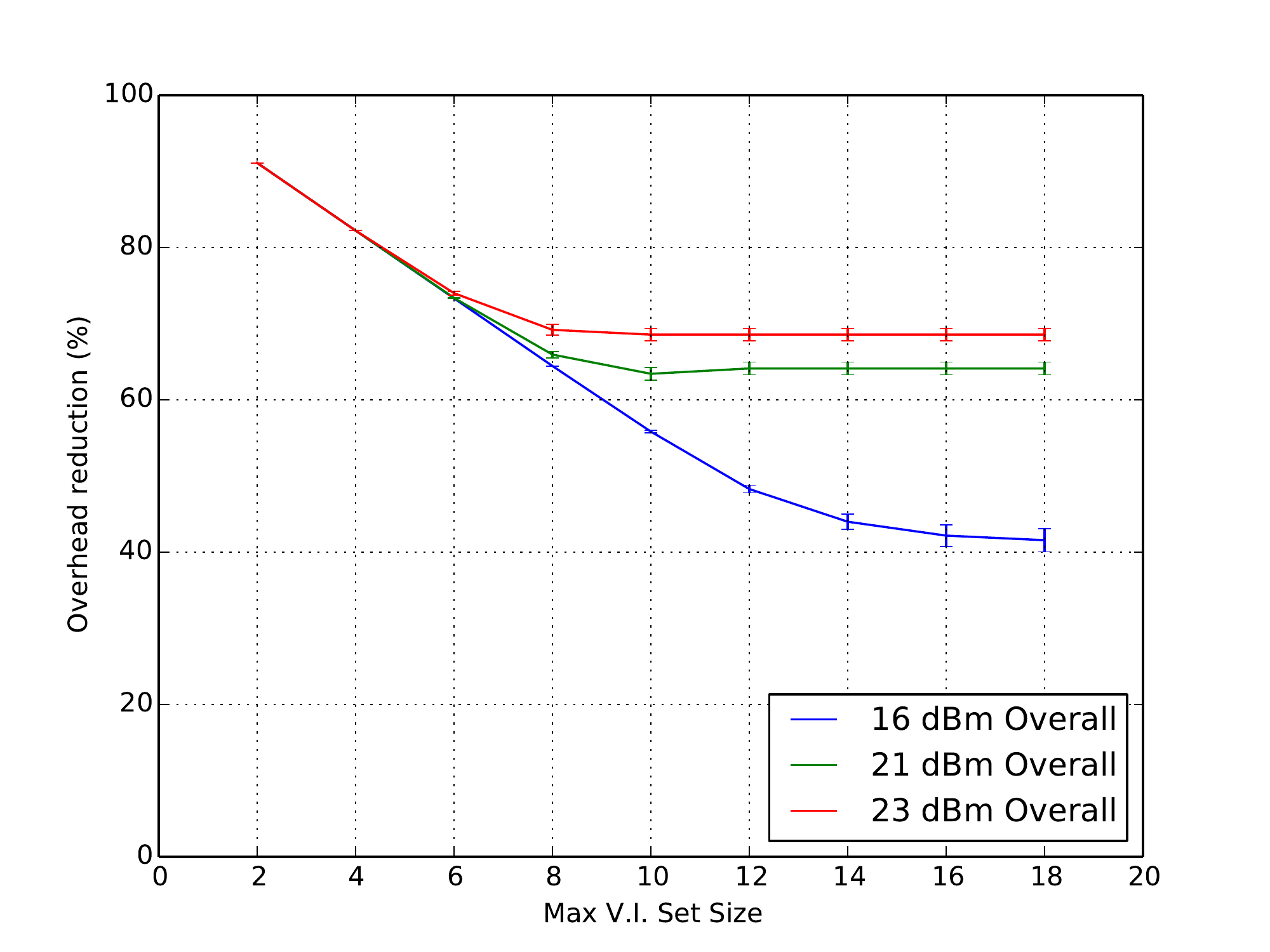}}           
        \caption{Performance evaluation of the algortithm for multi-technology information dissemination. The results for a given maximum number of virtual infrastructure ($k$) are  aggregated considering a number of requests of the simulation interval.  Error bars represent 95\% confidence intervals around the mean measure.}
        \label{fig:PERFeval}
\end{figure*}

Fig.~\ref{fig:PERFeval} presents the main results for the performance evaluation of the algorithm for multi-technology information dissemination. 
First, we analyze the impact of vehicle distribution and vehicle densities patterns on the algorithm performance.
For the three metrics presented in Fig.~\ref{fig:PERFeval} we can observe that the above mentioned parameters 
have little impact on the overall algorithm performance due to the small confidence intervals for each $k$.
As the transmit power increases the confidence intervals become smaller, which shows that the algorithm delivers better performance
in scenarios with improved node connectivity.

\begin{figure*}
        \centering
		\subfigure[\scriptsize 16~dBm]{\label{fig:del16}\includegraphics[width=0.32\textwidth]{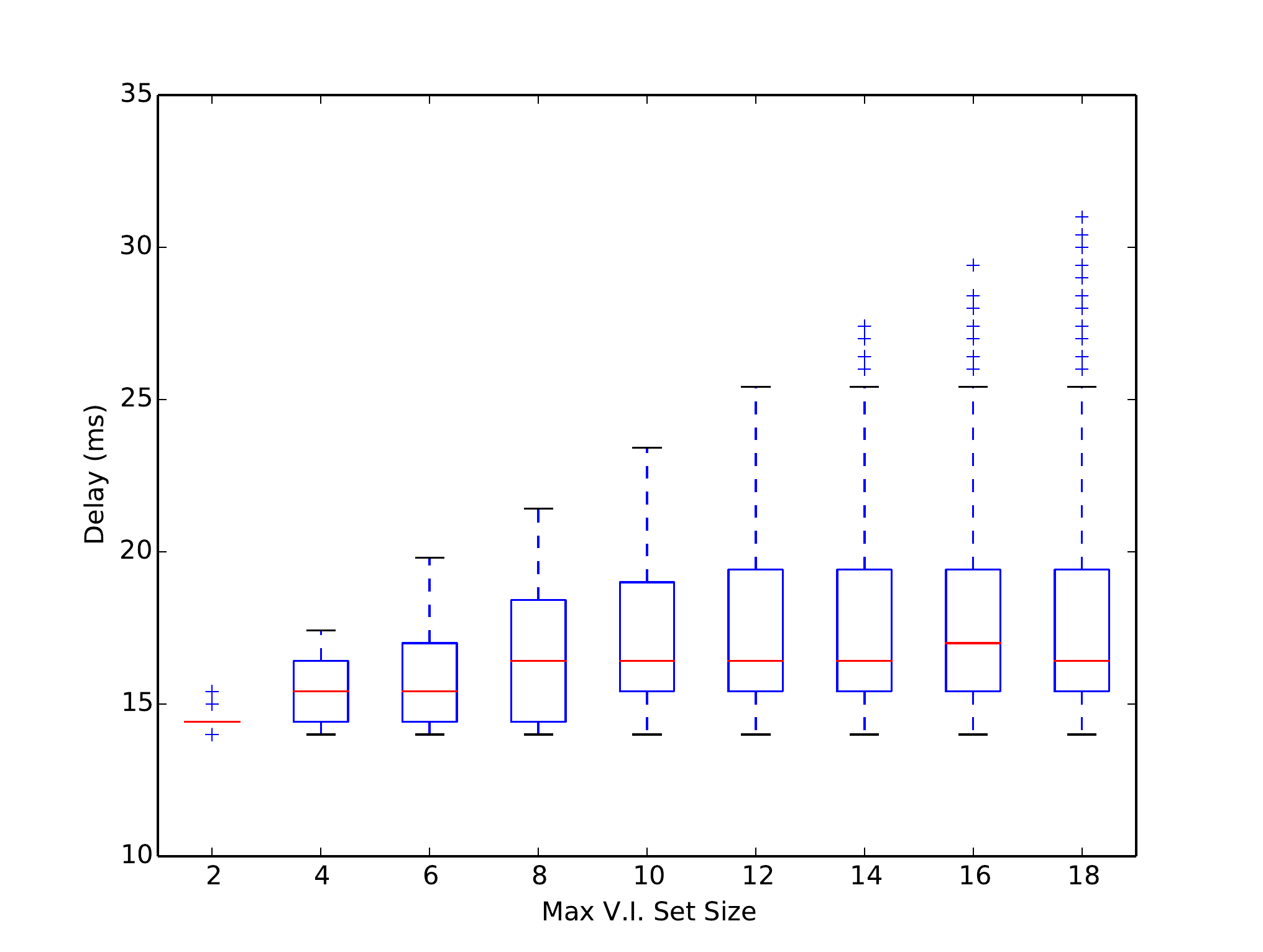}}
		\subfigure[\scriptsize 21~dBm]{\label{fig:del21}\includegraphics[width=0.32\textwidth]{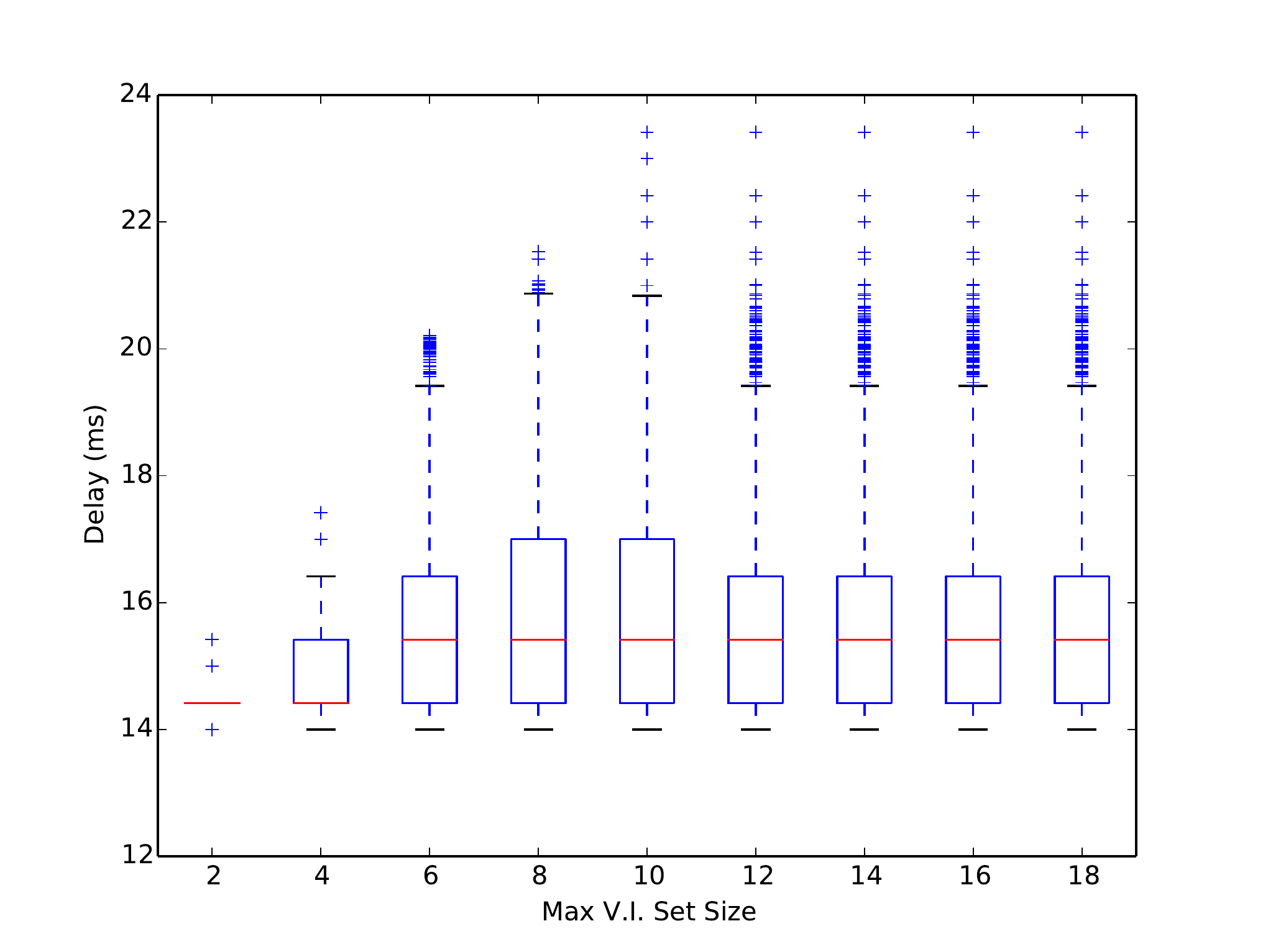}} 
		\subfigure[\scriptsize 23~dBm]{\label{fig:del23}\includegraphics[width=0.32\textwidth]{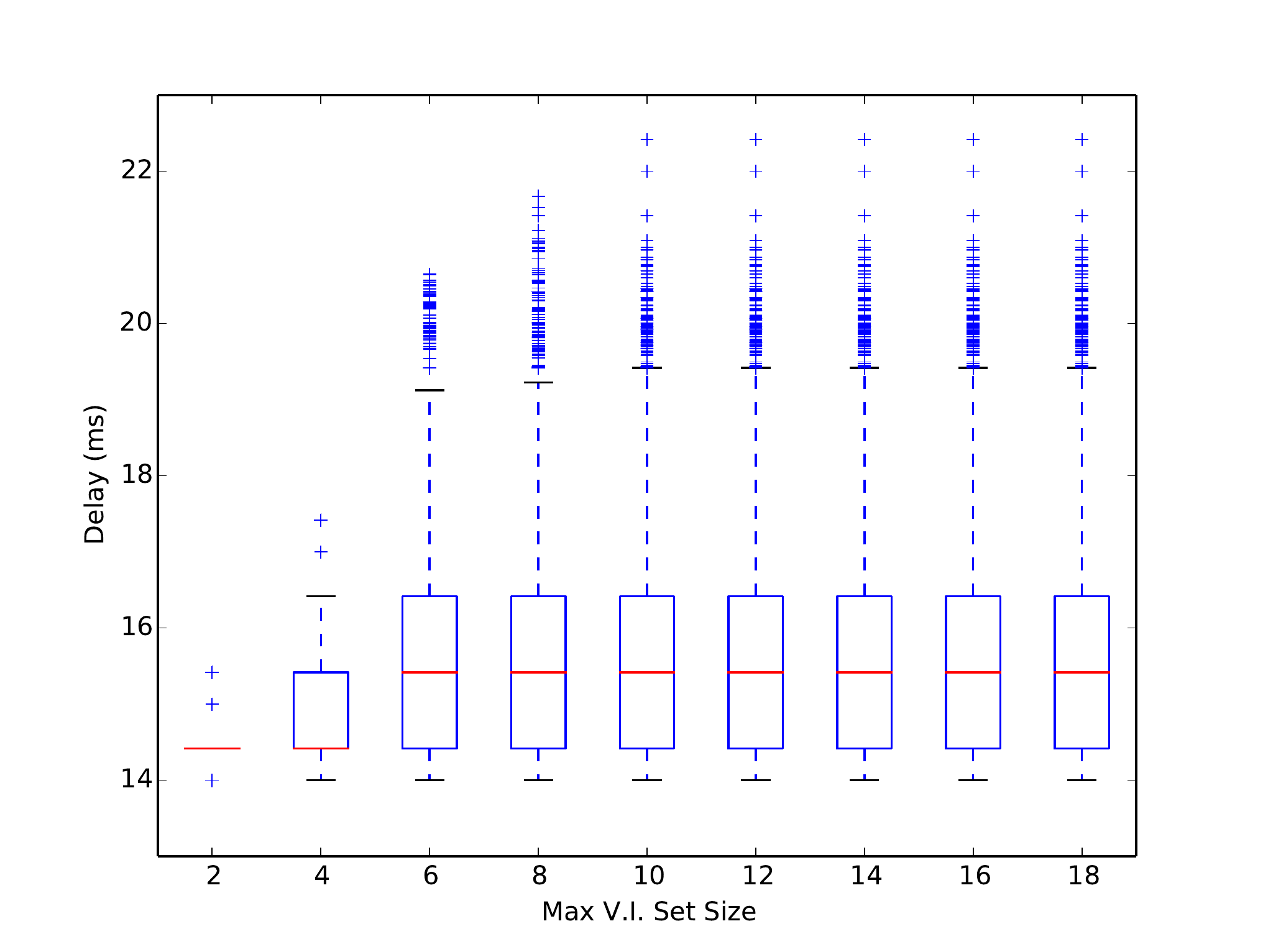}}           
        \caption{Delay performance of the algortithm for multi-technology information dissemination.}
        \label{fig:delayPERF}
\end{figure*}

It is also of interest to study the impact of controllable parameters, i.e. maximum allowed number of selected virtual infrastructure nodes ($k$) and transmit power,
on the overall system performance. Regarding the maximum number of selected virtual infrastructure ($k$), we can observe that this parameters has a considerable impact on the system performance.
As $k$ increases, more vehicles can be selected for disseminating information and consequently the covered area increases until 100~$\%$  as indicated in Fig.~\ref{fig:covVeh}. 
However, after a given threshold (variable for different transmit powers) increasing the maximum number of virtual infrastructure does not provide benefits in terms covered area.
From Fig. \ref{fig:chSize}, we can also conclude that the size of the virtual infrastructure set remains fairly constant after the above mentioned threshold is reached.
This occurs since the algorithm stops selecting virtual infrastructure after the stop criteria has been met.
From the results we should also highlight that for achieving a 100~$\%$ message penetration, the average virtual infrastructure set size is considerably small
when comparing with the total number of nodes in the dissemination area (e.g. 15$\%$ for a 23~dBm Tx power).
With respect to the overhead, the algorithm enables considerable reductions in terms of message exchange, specially for increasing transmission powers.
The overhead reductions arise mainly from the criterious node selection and offload from traffic from \ac{LTE} to ITS G5.
Fig.~\ref{fig:delayPERF} presents the end-to-end delay in message reception for different transmissions powers and $k$ values.
The results show that the algorithm delivers low latency values that are able to meet the requirements of the majority of applications.

With respect to the impact of transmission power, we can conclude that the increasing transmission powers improve the communication range 
and consequently lead to more stable node relations, which improves the algorithm performance. 
Increasing transmission power leads to faster convergence of the algorithm to maximum covered area
and a smaller size for the virtual infrastructure set, which provides bigger reductions in the overhead resource consumption. 
However, it should be highlighted that the algorithm performs well under more challenging propagation conditions.

%
%

\section{Conclusion} \label{sec:conclusions}
In this paper, we propose a centralized system for electing virtual infrastructure in a multi-technology vehicular environment based on the neighbor tables collected at vehicles.
The underlying mechanism is based on the computation of dissimilarity indexes between vehicles to improve information penetration  in a geographical area while considering
several constraints.

The presented results show the feasibility of the proposed system to achieve maximum message penetration with reduced overhead.
The system performance is influenced by the application requirements and by the transmission power.
The simulation results show that the algorithm selects a reduced set of vehicles as virtual infrastructure with minimal implications
in the communication performance in terms of delay but providing considerable overhead reductions.
Increasing the transmission power improves the performance and convergence rate of the algorithm.

As future work, we plan to compare the behavior of the system on different scenarios (urban, semi-urban and highway environment). 
As the current  implementation of the algorithm solely considers one-hop message distribution, we 
we intend to study the system performance multi-hop scenarios and multi-level virtual infrastructure.

\section*{Acknowledgment}
The authors would like to thank the financial support provided by the Mobility2.0 project, 
co-funded by the EU under the 7th Framework Programme for research (grant agreement no. 314129).


\bibliographystyle{IEEEtran}
\bibliography{draftIII_tex}

\end{document}